\documentclass[preprint, aps, pre, eqsecnum,amsmath,amssymb,showpacs]{revtex4}

\usepackage[final]{graphicx}

\newcommand{\Nabla}{\mbox{\bf\boldmath $\nabla$}}

\newcommand{\bean}{\begin{eqnarray}}
\newcommand{\eean}{\end{eqnarray}}
\newcommand{\bea}{\begin{eqnarray*}}
\newcommand{\eea}{\end{eqnarray*}}
\newcommand{\beq}{\begin{equation}}
\newcommand{\eeq}{\end{equation}}

\def\vereq#1#2{\lower3pt\vbox{\baselineskip1.5pt \lineskip1.5pt
\ialign{$\hfill##\hfil$\crcr#2\crcr\sim\crcr}}}
\def\gtrsim{\mathrel{\mathpalette\vereq>}}

\begin{document}

\title{\bf Spiral vortices traveling between two rotating defects in the
Taylor-Couette system}

\author{Ch.~Hoffmann, M.~L\"{u}cke, and A.~Pinter}
\affiliation{Institut f\"{u}r Theoretische Physik, \\ Universit\"{a}t
des Saarlandes, D-66041~Saarbr\"{u}cken, Germany\\}

\date{\today}

\begin{abstract}

Numerical calculations of vortex flows in Taylor-Couette systems with counter
rotating cylinders are presented. The full, time dependent
Navier-Stokes equations are solved with a combination of a finite difference
and a Galerkin method. Annular gaps of radius ratio $\eta=0.5$ and of several
heights are simulated. They are closed by nonrotating lids that produce
localized Ekman vortices in their vicinity and that prevent axial phase
propagation of spiral vortices. Existence and spatio temporal properties of
rotating defects, of modulated Ekman vortices, and of the spiral
vortex structures in the bulk are elucidated in quantitative detail.

\end{abstract}


\maketitle

\section{Introduction}
The spontaneous appearance of spiral vortices in the annular gap between the
concentric rotating cylinders of the Taylor-Couette system \cite{reviews} has
been stimulating research activities
\cite{DI84,B85,ZTL85,ALS86,GS86-GL88,LTKSG88,TESM89,E90,ETDS91,KP92,T94,CI94,AMS98,
SP00,HL00,MM02,CSBL02,PLH03,LPA04,HLP04} ever since their prediction \cite{KGD66}
and first observation \cite{S68}.
Spiral vortex structures bifurcate like the competing
toroidally closed Taylor vortices out of the rotationally
symmetric and axially homogeneous basic state of circular Couette flow (CCF),
albeit at different bifurcation thresholds \cite{LTKSG88,HLP04}.
The Taylor vortex flow (TVF) is rotationally symmetric and stationary while
the spiral vortex flow (SPI) breaks the rotational symmetry of the annular gap.
It oscillates globally in time by rotating azimuthally as a whole thereby
propagating axially.

The spiral pattern is effectively one dimensional like TVF. It is also
stationary when seen from a co-moving frame \cite{CI94}: the spiral fields do
not depend on time $t$,
axial coordinate $z$, and azimuthal angle $\varphi$ separately but only via the
combined phase variable $\phi= kz + M\varphi -\omega(k, M) t$. Here $k$ and $M$
are the axial and azimuthal wave numbers, respectively, and $\omega$ is the
frequency. In the $\varphi-z$ plane of an 'unrolled' cylindrical surface the
lines of constant phase $\phi$ are straight. An azimuthal wave
number $M>0$ implies a left handed spiral (L-SPI) while $M<0$ refer to
right handed spirals (R-SPI) with our convention of taking $k$ to be positive.
L-SPI and R-SPI being mirror images of each
other under the operation $z \to -z$ are symmetry degenerate flow states. Which
of them is realized in a particular experimental or numerical setup depends on
initial conditions and parameter history.

With the lines of constant phase in the $\varphi-z$ plane  being oriented for
both spiral types obliquely to the azimuthal 'wind' of the basic CCF both
spirals are advectively rotated by the latter like rigid objects.
The direction of the common angular velocity
$\dot{\varphi}_{SPI} = \omega(k,M)/M$ is the one of the inner cylinder's
rotation rate \cite{HLP04} which we take to be positive.
Due to the advection enforced rigid-body rotation of the spiral vortices
the phase of an L-SPI
($M>0$) is propagated axially upwards and that of an R-SPI ($M<0$) downwards.
Thus, the oscillatory flow structure of so called ribbons consisting of an
equal amplitude nonlinear combination of L-SPI and R-SPI rotates azimuthally but
does not propagate axially \cite{CI94}. On the other hand, the rotationally
symmetric ($M=0$) structure of toroidally
closed Taylor vortices is stationary: being parallel to the azimuthal CCF the
latter cannot advect these vortices.

Strictly speaking the axially homogeneous CCF and the TVF and SPI structures
exist with axially homogeneous amplitudes only in the theoretical idealizations
of axially unbounded or axially periodic systems. Translational symmetry breaking
conditions at the top and bottom end of the annulus generate (mostly local)
deviations in the basic state flow as well in the above mentioned vortex
structures. For example, the experimentally often used rigid non rotating lids
that close the annular gap enforce for any driving the well known stationary,
rotationally symmetric Ekman vortices
close to the lids \cite{PR81,GD82,ACDLH86,CSBL03,LPA04}. Their spatially varying wave
number and amplitude profile distinguishes them from the TVF
structure with axially homogeneous profiles.

In a sufficiently long system the
Ekman vortex structures close to the lids smoothly connect and transform to a
bulk TVF structure both patterns being stationary with common azimuthal
wave number $M=0$. So, then the question is: How do rotating and axially
propagating SPI vortices with $M\ne 0$ arise in the bulk when the
non propagating Ekman vortex structures being fixed at the lids prevent phase
propagation there? This is basically the problem that we elucidate here using
numerical simulations of the full 3D Navier-Stokes equations (NSE).
Surprisingly, it does not seem to have been addressed in such a detail in the
literature.

However, the influence of a finite system size on a traveling pattern like SPI
vortex flow has been explored, albeit from a more general point of view
\cite{BH79,ZTL85,ETDS91,KP92}.
Also the dramatic effects of nonrotating rigid lids on the flow in rather
short Taylor Couette systems has been investigated in detail for setups
where the vortex structures show strong axial variations \cite{CSBL02,LPA03}.

Our paper is organized as follows:
In Sec.~\ref{SEC:System} we introduce the notation, the control parameters, the
basic equations, and the method used to simulate the Taylor Couette system.
Section \ref{SEC:Results} contains our results concerning the transient dynamics
of spiral generation, the steady state structure and dynamics in particular of
the rotating defects, and the stability of SPI flow. The last section contains a
conclusion.

\section{System and theoretical description}
\label{SEC:System}

We present numerical results for the vortex flow in Taylor-Couette
systems with counter-rotating cylinders. The radius ratio $r_1/r_2$ of inner to
outer cylinder is $\eta=0.5$. Various aspect ratios $\Gamma=L/d$ of cylinder
length $L$ to gapwidth $d=r_2-r_1$ are considered in the range
$5\le \Gamma \le 16$. The fluid in the annulus is taken to be
isothermal and incompressible with kinematic viscosity $\nu$.
To characterize the driving of the system, we use the Reynolds numbers
\begin{eqnarray}
R_1=r_1\Omega_1 d/\nu \,\, ; R_2=r_2\Omega_2 d/\nu \, .
\end{eqnarray}
They are just the reduced azimuthal velocities of the fluid at the inner and
outer cylinder, respectively, where $\Omega_1$ and $\Omega_2$ are the
respective angular velocities of the cylinders. The inner one is always
rotating counterclockwise so that $\Omega_1$ and $R_1$ are positive.

Throughout this paper we measure lengths in units of the gapwidth $d$. The
momentum diffusion time $d^2/ \nu$ radially across the gap is taken as the
time unit. Thus, velocities are reduced by $\nu /d$.
With this scaling, the NSE take the form
\begin{eqnarray}
\partial_t {\bf u} = \Nabla^2 {\bf u} -
({\bf u}\cdot \Nabla){\bf u} - \Nabla p \,.
\end{eqnarray}
Here $p$ denotes the pressure reduced by $\rho \nu^2/d^2$ and $\rho$ is the mass
density of the fluid. Using cylindrical coordinates, the velocity field
\begin{eqnarray}
{\bf u}=u\,{\bf e}_r + v\,{\bf e}_\varphi + w\,{\bf e}_z
\end{eqnarray}
is decomposed into a radial component $u$, an azimuthal one $v$, and an
axial one $w$.

The NSE were solved numerically with a finite differences method
in the $r-z$ plane combined with a spectral decomposition in $\varphi$
\begin{eqnarray}\label{EQ-expansion}
 f(r,\varphi,z,t)=
 \sum_{m=-m_{max}}^{m_{max}} f_m(r,z,t)\,e^{im\varphi} \, .
\end{eqnarray}
Here $f$ denotes one of $\{u,v,w,p\}$ and $m_{max}=8$ was chosen
for an adequate accuracy. To simulate annuli that are bounded by stationary
lids at $z=0$ and $z=\Gamma$ we imposed there no-slip boundary conditions.

The calculations were done on homogeneous staggered grids with common
discretization lengths $\Delta r=\Delta
z=0.05$ which have shown to be more accurate than non-homogeneous grids.
Time steps were always well below the von
Neumann stability criterion and by more than a factor of three below the
Courant-Friederichs-Lewy criterion (cf. \cite{HLP04} for details of the
numerical calculations). From various control calculations done with different
$m_{max}$ and/or the grid spacing we conservatively conclude that typical SPI
frequencies have an error of less than about 0.2\% and that typical velocity
field amplitudes can be off by about 3 - 4\%. Furthermore, good agreement with
experimental spirals was found --- cf. Figures 8 and 9 of \cite{HLP04}.

For diagnostic purposes we also evaluated the complex mode amplitudes
$f_{m,n}(r,t)$ obtained from a Fourier decomposition in axial direction
\begin{eqnarray}\label{FFT-expansion}
 f_m(r,z,t)=
 \sum_{n} f_{m,n}(r,t)\,e^{in(2\pi/\Gamma)z} \, .
\end{eqnarray}
Note that $m$ is the index of a particular azimuthal mode occurring in the
representations (\ref{EQ-expansion}) and (\ref{FFT-expansion}) while we use $M$ to
identify the azimuthal wave number of a particular solution. So, for example, a
$M=-1$ flow state is a R-SPI with azimuthal wave number $M=-1$ that will contain
in general several $m$ modes.

\section{Results} \label{SEC:Results}
\begin{figure}[ht]
\begin{center}
\includegraphics[width=12cm]{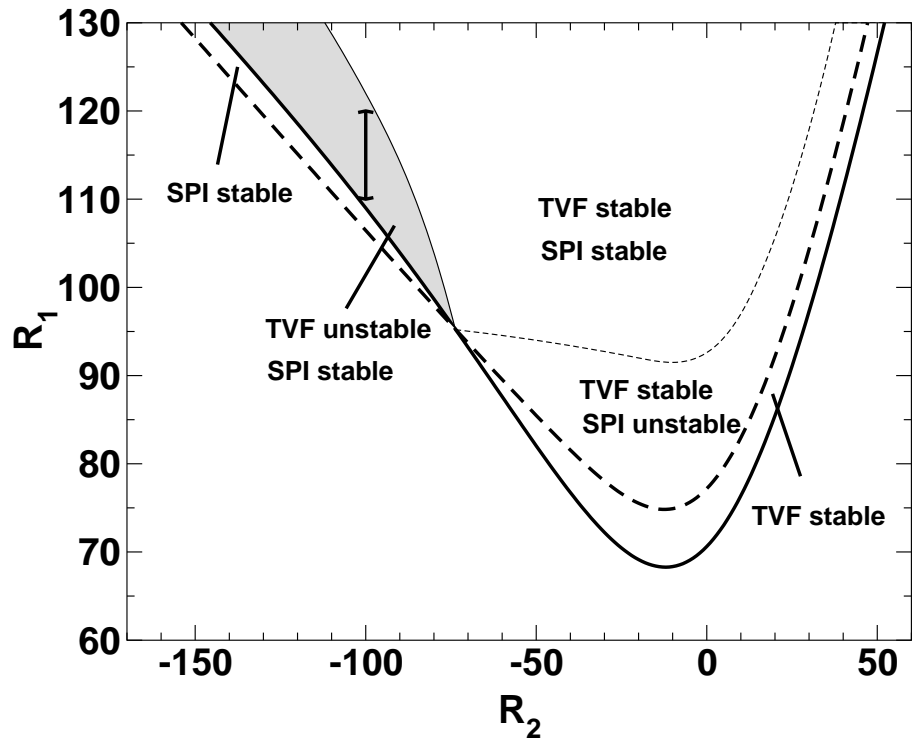}
\end{center}
\caption{Phase and stability diagram of TVF($M=0$) and SPI($M=\pm 1$) solutions
subject to axially periodic boundary conditions imposing the axial wavelength
$\lambda=1.6$. Thick full (dashed) line denotes the bifurcation threshold for
the TVF (SPI) solution out of CCF. Vertical bar indicates the range of $R_1$
values for which simulations of finite length systems with rigid stationary lids
are presented here.}
\label{fig1}
\end{figure}

For our finite-length annuli with stationary lids at their ends we kept the
outer cylinder rotation Reynolds number fixed at $R_2=-100$. Results were
obtained for $R_1$ in the range $110\le R_1 \le 120$ that is marked by a
vertical bar in Fig.~\ref{fig1}.

This figure shows for reference purposes the phase and
stability diagram of TVF($M=0$) and SPI($M=\pm 1$) solutions subject to axially
periodic boundary conditions. The range $110\le R_1 \le 120$ to be explored here
lies in a control parameter region where both, SPI and TVF solutions exist with
the former (latter) being stable (unstable) under periodic boundary conditions.
The bifurcation thresholds out of the CCF lie at $R_1=106.5$ for SPI and at
$R_1=108.9$ for TVF.

Strictly speaking these axially periodic solutions do not exist in systems of
finite axial length that are bounded by rigid lids: Ekman vortices
\cite{PR81,GD82,ACDLH86,CSBL03,LPA04}
always appear already subcritically near the lids with a spatially varying wave
number and amplitude profile that distinguishes them from the homogeneous TVF
structure. Also SPI flow can be realized with constant amplitude and wave number
only in the bulk at sufficiently large distance from the lids.

\subsection{Transient dynamics of spiral generation in the bulk}

Here we want to show how spirals occur in the bulk of a $\Gamma=12$ system as a
representative example of commonly used set-ups in experiments. We start from
rest --- to be precise from the quiescent fluid plus infinitesimal white noise
in all velocity fields. Then the rotation rates of the cylinders are stepped up
instantaneously to supercritical final values of $R_1$ and $R_2$ for which SPI
flow is stable and TVF is unstable under axially periodic boundary conditions,
cf. Fig.~\ref{fig1}. Step up from a subcritical driving entails a similar
transient.

\subsubsection{Front propagation of unstable TVF into unstable CCF}

Figs.~\ref{fig2} and~\ref{fig3} show the longterm evolution of the flow for
the case of $R_1=110$ which lies about 1\% (2\%) above the TVF (SPI) threshold.
However, first, the unstable CCF flow is growing radially in the bulk and simultaneously
the Ekman vortices are growing near the lids \cite{LMW85}. Both occurs
on a fast time scale of about 1-2 radial diffusion times which are not resolved
in Figs.~\ref{fig2} and~\ref{fig3}. Then TVF fronts are
propagating axially into the bulk from the Ekman vortex structures near the
lids \cite{LMW85} --- note that $M=0$ TVF can grow at supercritical driving
independent of
its stability behavior. So here we have a front of an unstable structured state
that propagates into an unstable unstructured one. The velocity of the TVF
fronts is rather
large progressing at least 5 gapwidths per unit diffusion time. So after about
5 diffusion times the fully developed unstable TVF is established with
homogeneous amplitude and wave number profile in the bulk in
equilibrium with the axially varying Ekman vortex structures near the lids,
cf. row A of Fig.~\ref{fig2}. This TVF growth scenario is dominated by
the large deterministic forces that drive Ekman vortex flow near the lids and
thus is largely insensitive to the small initial noise.

\subsubsection{Transformation of unstable TVF into stable SPI flow}

\begin{figure}[ht]
\begin{center}
\includegraphics[width=16cm]{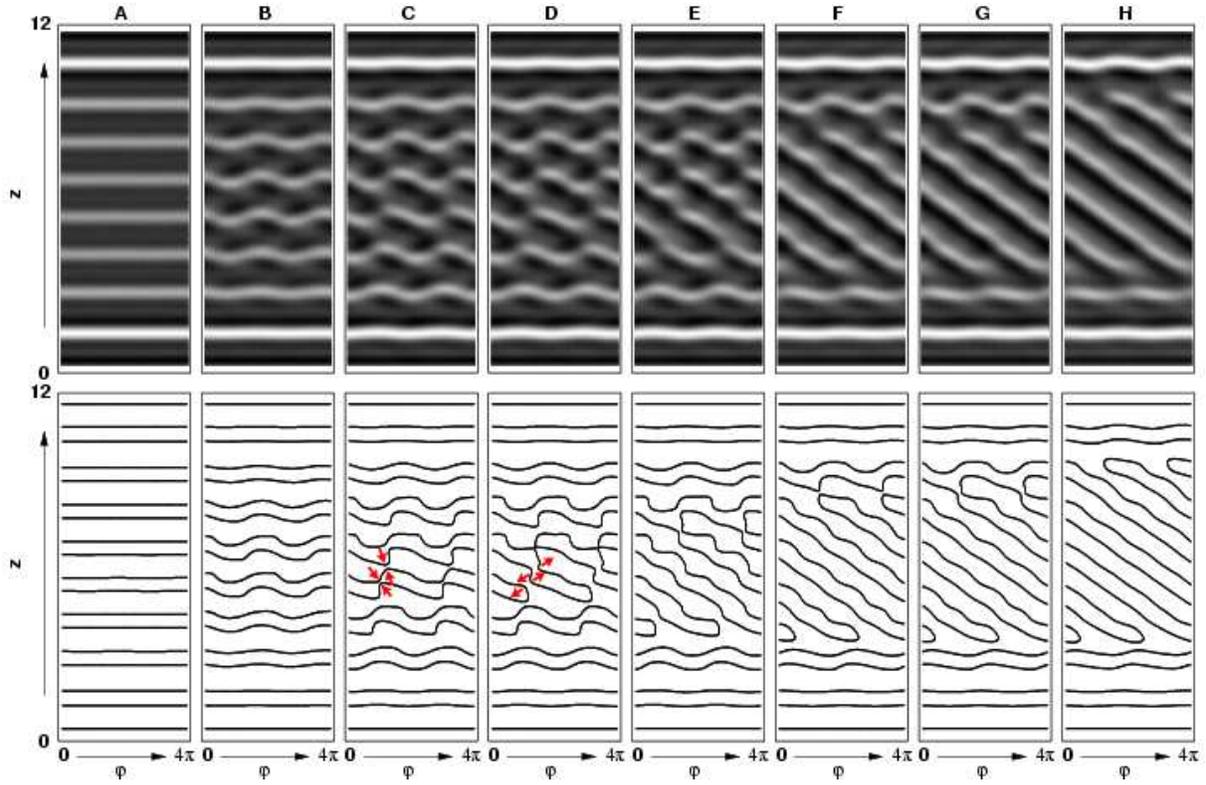}
\end{center}
\caption{Snapshots of the time evolution towards L-SPI flow in the bulk. The
snapshot times A-H are indicated in Fig.~\ref{fig3}. Top row shows the radial
velocity field $u$ in an unrolled cylindrical $\varphi$-$z$-surface
at mid gap by gray scale plots with white (black) denoting radial out (in)
flow. Bottom row shows the node positions of $u$. For better visibility the
plots are periodically extended to an azimuthal interval of $4\pi$. The initial
condition is the
fluid at rest plus infinitesimal white noise in all velocity fields. Final
Reynolds numbers are $R_1=110, R_2=-100$. Aspect ratio $\Gamma$=12.
}
\label{fig2}
\end{figure}

\begin{figure}[ht]
\begin{center}
\includegraphics[width=15cm]{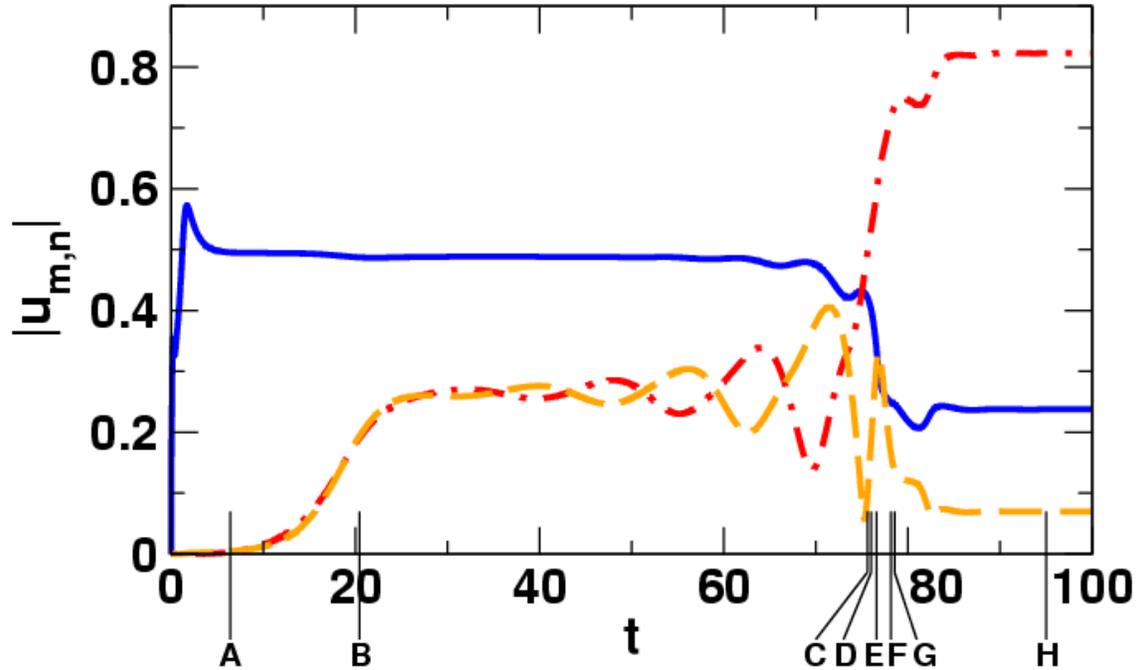}
\end{center}
\caption{Time evolution of the dominant mode amplitudes $|u_{m,n}|$
(\ref{FFT-expansion}) of the radial velocity field at mid gap. Shown is the
transient towards L-SPI flow in the bulk that is documented in Fig.~\ref{fig2}
by snapshots. Full line: TVF mode $m=0$. Dashed dotted line: L-SPI mode $m=1$.
Dashed line: R-SPI mode $m=-1$. For the aspect ratio $\Gamma$=12 considered here
the dominant axial mode index in the decomposition (\ref{FFT-expansion}) is
$n=\pm8$.}
\label{fig3}
\end{figure}

Starting with this TVF configuration, we illustrate in Fig.~\ref{fig2} the
further time
evolution of the vortex flow. To that end we show in the top row snapshots of
the radial velocity field $u$ in an unrolled cylindrical $\varphi$-$z$-surface
(that is azimuthally extended to $4 \pi$ for better visualization)
by gray scale plots. The bottom row contains snapshots of the node
positions of $u$ at mid gap. These snapshots cover a time
interval of about 100 radial diffusion times. The snapshot times are marked in
Fig.~\ref{fig3} which exhibits the dynamics of the dominant characteristic
mode amplitudes for TVF ($M=0$) and SPI ($M=\pm1$), respectively.

Snapshot (A) in Fig.~\ref{fig2} shows that by this time the rotational symmetric
TVF state has been established in the
bulk. The Ekman vortices of higher flow intensity are marked by the
brightest outflow line near each lid. By the time B the $m=\pm1$ modes that
break the rotational symmetry have grown sufficiently to see the wavy
deformation of the still dominant $M=0$ TVF in snapshot (B). Here the
amplitudes of $m=1$ and $m=-1$ modes are still of equal size giving rise to
an azimuthally rotating modulation of the TVF almost harmonic behavior. Then
the amplitudes of the $m=1$ and $m=-1$ modes start to oscillate in counterphase
with growing oscillation amplitude while
the $m=0$ mode does not change much, cf. Fig.~\ref{fig3}. But shortly before
time C the $m=1$ L-SPI mode takes off: it continues to increase
while the $m=-1$ R-SPI mode and also the $m=0$ mode decrease.

This mode behavior reflects the fact that starting in the bulk the TVF vortices
become more and more deformed. The nodes of $u$ in the bottom row of
Fig.~\ref{fig2} show how the vortices approach each other (cf. arrows in C) and
get pinched together at a defect that "cuts" them into two. They move apart
(cf. arrows in D), get tilted in the $\varphi$-$z$ plane of Fig.~\ref{fig2}, and
reconnect differently to form locally a spiral vortex pair. This
defect formation and reconnection is repeated at two new locations further
upwards and downwards towards the lids. The defect propagation is stopped by
the strong Ekman vortex structures. They are only slightly indented by the
rotating defect in the final state.

So, in the final state at time H the bulk is filled with an axially upwards
propagating L-SPI structure. Its phase is generated by a defect that is
rotating in the lower part of the system. The spiral phase is annihilated at
another rotating defect in the upper part of the system.

That here the $m=1$
mode wins the mode competition leading finally to a L-SPI structure in the bulk
while the $m=-1$ mode gets suppressed is not due to an intrinsic selection
mechanism. It merely reflects the fact that in this particular transient the
initial white noise condition of the velocity field had a slightly higher
content of L-SPI modes. In other runs with another noise realization the
R-SPI could equally well win the competition given that our random number
generator for producing the white noise is unbiased.

\subsection{Steady state structure and dynamics}

By the time H in Fig.~\ref{fig2} transients have died out and the flow has
reached its final state. It consists of an L-SPI structure in the bulk with
azimuthal wave number $M=1$ (i.e., one pair of spiral vortices),
slightly modulated Ekman vortex structures that are localized next to the two
lids, and two rotating but axially not propagating defects. This flow
structure is rotating as
a whole like a rigid body with a global rotation rate $\omega$ into the same
positive $\varphi$-direction as the inner cylinder. However, the spiral
rotation rate $\omega$ is somewhat smaller than the one of the inner cylinder
\cite{HLP04}. Driven by this rotation the L-SPI phase in the bulk is
propagating axially upwards.

We should like to stress that the flow in Fig.~\ref{fig2}H contains in the
decomposition (\ref{FFT-expansion}) besides the dominant  $m=1$ SPI modes not
only $m=0$ modes that are related primarily to the Ekman vortex structures
but also a significant $m=-1$ contribution, cf. Fig.~\ref{fig3}. The rotating
defects and the rotating modulations of the Ekman vortices are the reason for
the presence of $m=-1$ modes in addition to $m=1$ modes. In fact, locally, in
the axially non propagating flow regions of the rotating defects and of the
rotating Ekman vortex modulations they combine to axially standing oscillations.

\subsubsection{Structure of the rotating defects} \label{SEC:rot-defect}

The Ekman vortices near the lids do not propagate but remain spatially localized
while the SPI vortices propagate. The connection between
these topologically different vortex structures is provided by a pair of rotating
defects: The one close to the lower Ekman vortex structure generates the L-SPI
phase where two lines of nodes of the SPI $u$ field appear in Fig.~\ref{fig2}H
in the form of a U tilted to the left. The defect close to the upper Ekman 
vortex structure annihilates the phase when the lines of the SPI nodes join
again. With the two defects locating the beginning and end, respectively, of 
the spiral vortex pair the former may be seen as pinning the latter.

\begin{figure}[ht]
\begin{center}
\includegraphics[width=14cm,angle=0]{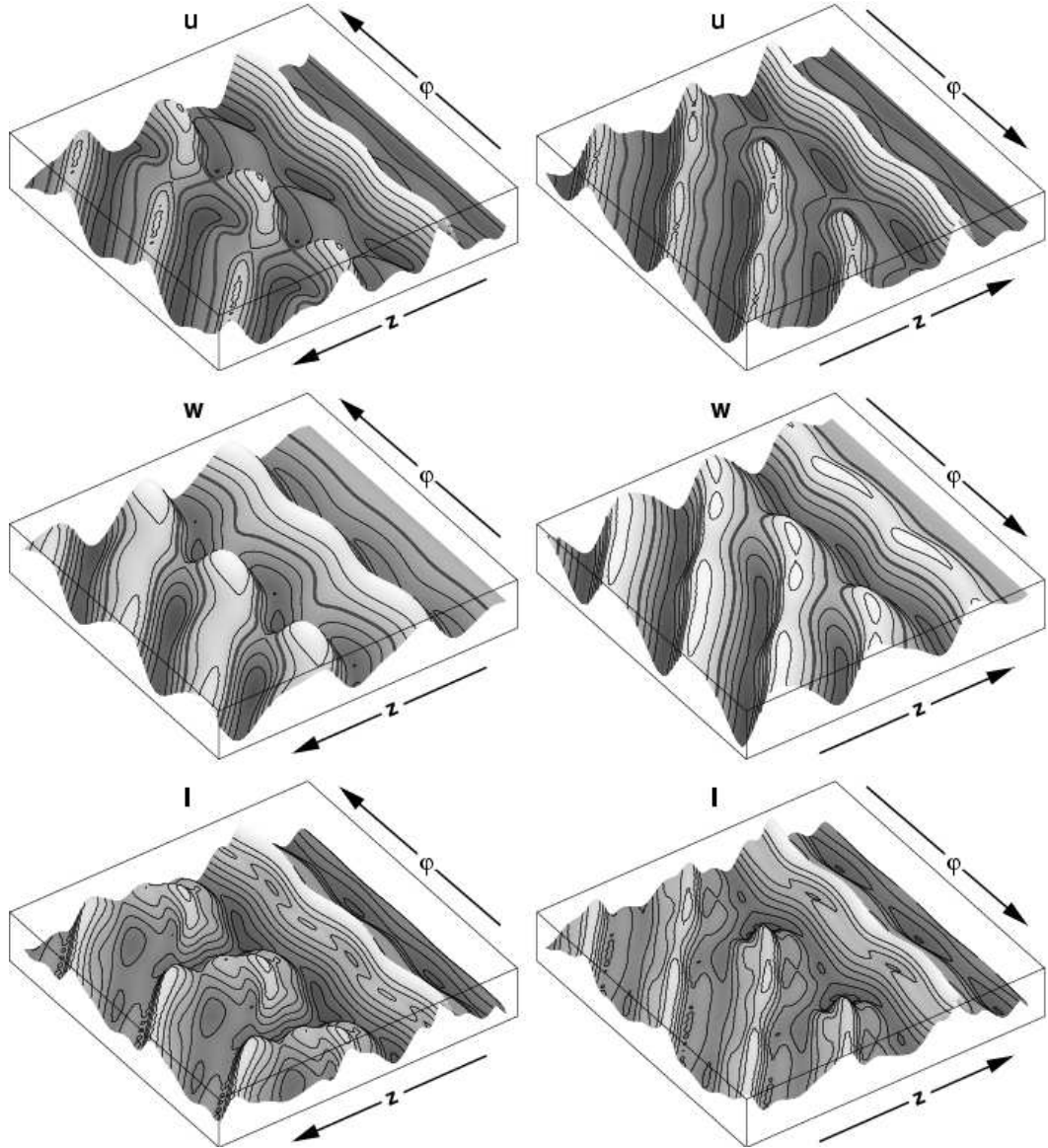}
\end{center}
\caption{Flow structure in the vicinity of the rotating defects. The left (right)
column documents the L-SPI generation (annihilation) near the
lower (upper) Ekman vortex structure. Shown are the fields $u, w$, and
$I=\sqrt{u^2 + w^2}$ at mid gap over the $\varphi-z$-plane. Note that there is
only one defect at top and bottom: for better
visibility the plots are periodically extended to an azimuthal interval of
$4\pi$. Thick (red) isolines in the  $u, w$ plots show the node positions of
these fields. Parameters are $R_1=115, R_2=-100, \Gamma$=12.}
\label{fig4}
\end{figure}

The flow structure in the
vicinity of the two rotating defects is shown in Fig.~\ref{fig4} for $R_1=115$.
The gray scale plots show from top to bottom $u,w$, and the intensity
$I=\sqrt{u^2 + w^2}$ over the $\varphi-z$-plane. The left (right) column
documents the L-SPI generation (annihilation) near the
lower (upper) Ekman vortex structure. The zeroes of $u$ and $w$ are shown by
thick red lines. Their U-turn marks the location of the defects. 
The phase generating defect in the three fields of the left column that 
disrupts the bottom Ekman vortex structure has a
slightly more complex structure than the phase annihilating defect in the right
column. One sees that the Ekman vortices closest
to the lids are modulated by the rotating defect but otherwise remain intact.
Figs.~\ref{fig2}H and ~\ref{fig4} show also that the upwards
propagating spiral vortices compress the Ekman vortex structure near the upper
lid and dilate the one near the lower lid. Thus, the upwards traveling SPI
phase "pushes" the Ekman vortices towards the top lid and "pulls" them away
from the bottom lid.

\begin{figure}[ht]
\begin{center}
\includegraphics[width=8cm]{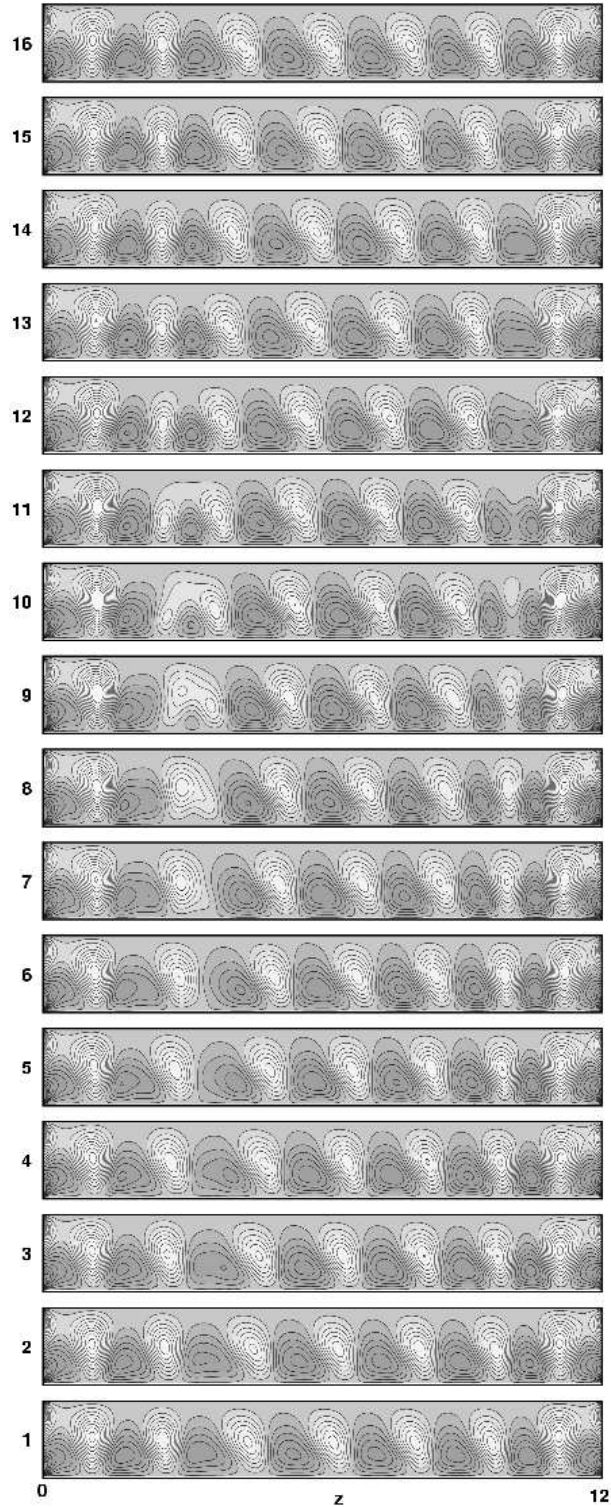}
\end{center}
\caption{
Steady state L-SPI generation, propagation, and annihilation covering one period. 
Shown are color-coded snapshots of the azimuthal vortex flow field $v-v_{CCF}$
together with its isolines in the $z-r-$plane. They are taken 
at fixed $\varphi$ at times $t_n\omega = 2\pi n/16$ or,
equivalently, at fixed $t$ at azimuthal angles $\varphi_n =2\pi n/16$ for $n$=1
to 16 as indicated. The inner (outer) cylinder is located at the bottom (top) 
of each snapshot. Parameters are $R_1=110, R_2=-100, \Gamma$=12 as in 
Figs.~\ref{fig2} and \ref{fig3}.
}
\label{Fig:v-vCCF-movie}
\end{figure}

Fig.~\ref{Fig:v-vCCF-movie} shows in more detail the spatiotemporal dynamics of
generation, propagation, and annihilation of SPI vortices over one period. To
that end snapshots of the azimuthal vortex flow field $v-v_{CCF}$ are taken in
the $z-r-$plane at fixed $\varphi$ at times $t_n\omega = 2\pi n/16$ or,
equivalently, at fixed $t$ at azimuthal angles $\varphi_n =2\pi n/16$. In
snapshots 1-8 the fourth vortex from the left, $z$=0, expands. Then, at $n$=8-9
a new one starts to grow close to the inner cylinder thereby marking the
defect. In snapshots 10-14 this new vortex continues to grow and to expand
towards the outer cylinder. Simultaneously, at $n$=9-11 the old fourth vortex
splits into two single vortices with the same direction of rotation --- 
one to the left and one to the right of the new one. The
right neighbor is displaced upwards and propagates away. Vortex annihilation
proceeds by squeezing the fourth vortex from the top, say, at $n$=9-10 and by
merging its two neighbors in snapshots 10-12. 

For $R_1=110$ (Fig.~\ref{fig2}H),
i.e., close to the SPI bifurcation threshold the axial extension of the spiral
region is not as large as, say, for $R_1=115$ (Fig~\ref{fig4}). In fact, in the
range $110\le R_1 \le 117$ the bulk SPI region increases with increasing $R_1$
by displacing the Ekman vortex structures as the SPI amplitudes grow.
Even stronger rotation speeds $R_1>117$, however,
seem to prefer TVF: spirals are more and more displaced out of the
boundary region.

\subsubsection{SPI versus TVF modes}

\begin{figure}[ht]
\begin{center}
\includegraphics[width=8cm]{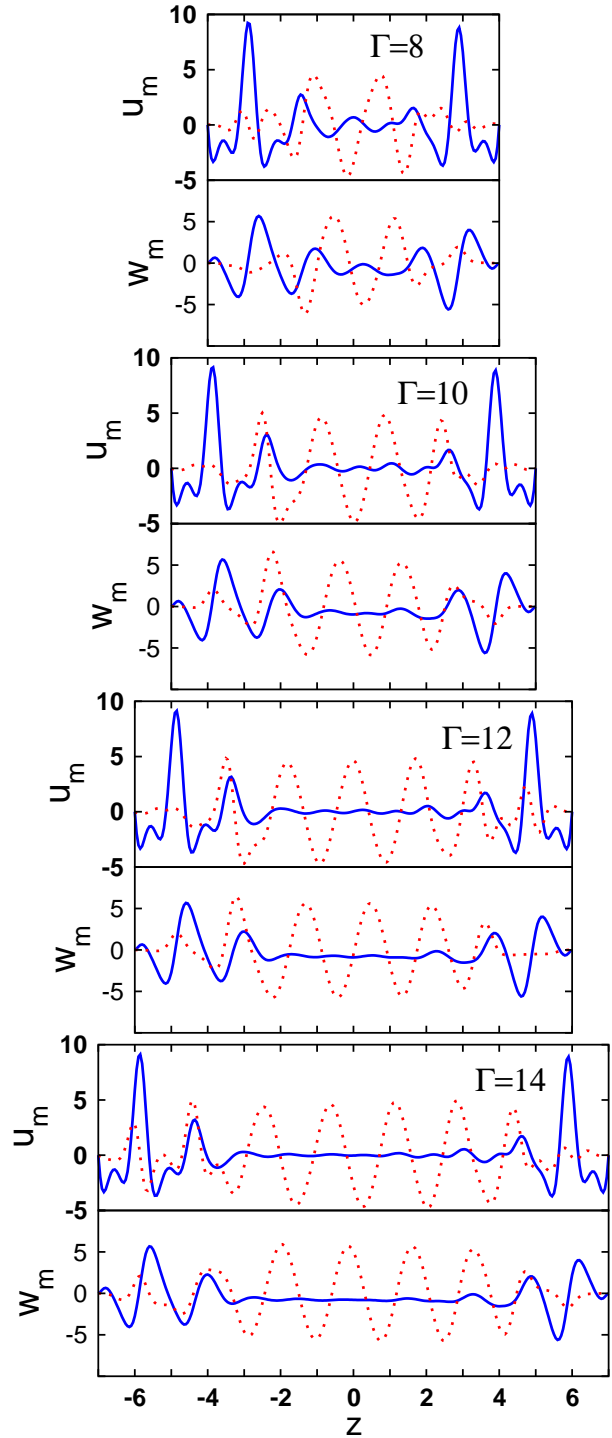}
\end{center}
\caption{Axial profiles of the dominant contributions in the decomposition
(\ref{EQ-expansion}) of the velocity fields from TVF and SPI modes. Full blue
(dotted red) line show snapshots of the real parts of $m=0$ TVF ($m=1$ SPI)
Fourier modes of $u$ and $w$ at mid gap in systems of different length $\Gamma$.
Parameters are $R_1=115, R_2=-100$.}
\label{fig5}
\end{figure}

In Fig.~\ref{fig5} we show axial profiles of the dominant contributions in the
decomposition
(\ref{EQ-expansion}) of the velocity fields from TVF and SPI modes. Full blue
(dotted red) line show snapshots of the real parts of $m=0$ TVF ($m=1$ SPI)
Fourier modes of $u$ and $w$ at mid gap in systems of different length $\Gamma$.

One sees that for the fixed $R_1=115$ shown in Fig.~\ref{fig5}
the extension of the $m=0$ Ekman vortex systems into the bulk
and their structure remain unchanged when $\Gamma$ is changed. However, at
$\Gamma \simeq 10$ the tails of the exponentially decreasing Ekman vortex flow
created by the two lid start to visibly overlap in the bulk.

On the other hand, the axial extension of the SPI vortex structure (dotted red
lines) in the bulk adjusts itself to the cylinder length.
The amplitude of the $m=1$ SPI mode is constant in the bulk
and it decays exponentially towards the lids. But it reaches well into the
Ekman vortex dominated region. This behavior reflects the rotating modulation
of the Ekman vortices that is caused by the rotating defect between SPI and
Ekman vortices. To sum all this up: decreasing the cylinder length shrinks the
bulk range where spirals exist.

\subsubsection{SPI wave number and frequency selection}

\begin{figure}[ht]
\begin{center}
\includegraphics[width=12cm]{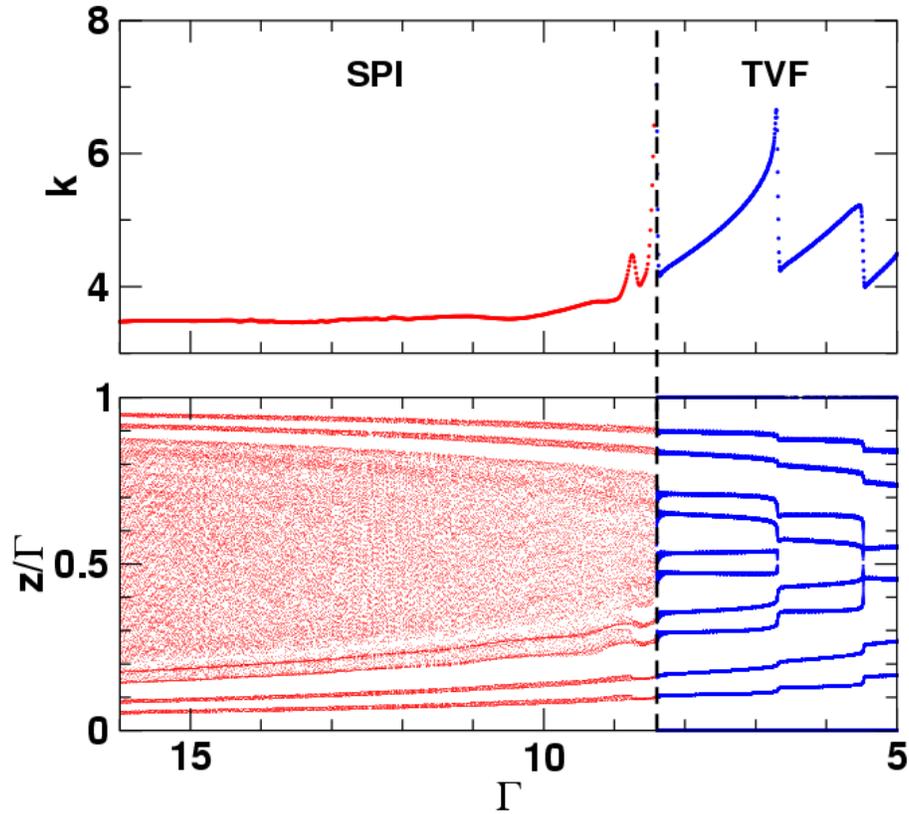}
\end{center}
\caption{Spatio temporal properties of the vortex flow as a function of system
length. $\Gamma$ was ramped down in steps of 0.05 by moving
the top lid downwards with about 2 radial diffusion times between successive
steps. For each $\Gamma$ the nodes of $u$ were monitored during this time
interval at discrete times at a fixed $\varphi$. Bottom plot shows the axial
distribution of these nodes by dots, cf. text for further information and
discussion. Top plot shows the wave number in the vicinity of mid height
location $z=\Gamma/2$ together with error bars explained in the text. The
vertical dashed line marks the transition from SPI
in the center part to TVF in this ramp 'experiment'. Parameters are
$R_1=115, R_2=-100$.}
\label{fig6}
\end{figure}

Fig.~\ref{fig5} indicates that the SPI structure at mid height (that is defined
in Fig.~\ref{fig5} to lie at $z=0$ for presentation reasons)
is the same, irrespective of the length of the system, over a wide range of
$\Gamma$. The observation of such a unique selection of the SPI structure is
corroborated by the fact that the SPI wave number $k$ measured in the vicinity
of the mid height position is practically independent of $\Gamma$, cf. top plot
of Fig.~\ref{fig6}. The selected SPI wave number
varies between $k=3.47$, $\lambda=2\pi/k=1.81$ ($\Gamma=16$) and $k=3.57$,
$\lambda=1.76$ ($\Gamma=10$).

Here it is worth mentioning that the corresponding SPI wavelength of
$\lambda \simeq 1.76$ has been observed in experiments \cite{SP00} done
in a system of length $\Gamma=12$. Furthermore, also the numerically
determined SPI flow structure agrees almost perfectly with the one obtained by
the afore mentioned laser-Doppler velocimetry measurements, cf. Fig.~8 of
Ref.~\cite{HLP04}. The selected frequency is $\omega \simeq 30.3$
so that the SPI phase
propagates axially with phase velocity $\omega /k \simeq 8.6$.

Fig.~\ref{fig6} shows results of a numerical simulation
in which the length $\Gamma$ was ramped down from $\Gamma=16$ to $\Gamma=5$ in
steps of $\Delta \Gamma=0.05$ by moving the top lid downwards. The time
intervals
between successive steps were about 2 radial diffusion times so that the SPI
phase had always enough time to propagate from one end to the other.

In the bottom plot of Fig.~\ref{fig6} we show for each $\Gamma$ the
axial distribution of the nodes of $u$ by dots. The nodes were
monitored at discrete times during this time interval at a fixed
$\varphi$. So, for example, the broadened lines near the top and
bottom lids denote the narrow axial excursions of the locations of
the Ekman vortices being modulated by the rotating defects. On the
other hand, the homogeneously distributed dots in the center reflect
the propagating SPI phase. The errorbars in the top plot come from
(i) the finite sampling rate which in general is not commensurate
with the time period of the propagating structure and (ii)  from the
fact that the nodes of $u$ which are used to measure the wavelength
lie (depending on that incommensurability) somewhere in a region 
around mid-height.

We observed the same SPI frequency and wave number selection also in
the upwards ramp described in Sec.~\ref{SEC:decreaseGamma}. Starting
with TVF at small $\Gamma$ the SPI appeared there only at $\Gamma
\simeq 10$. So, whenever SPI flow was realized in a substantial part
of the system with homogeneous amplitude then its frequency and wave
number was uniquely selected within our numerical accuracy.

\subsection{Stability of SPI flow}

\subsubsection{Decreasing $\Gamma$}

When in the above described ramping 'experiment' the length has fallen below
$\Gamma \simeq 8.3$ the system has become too small to allow for a
propagating SPI phase in the center. Instead stationary $M$=0 Ekman and TVF is
realized throughout the system with 10 nodes in the bulk, cf., right part of
Fig.~\ref{fig6}. Reducing $\Gamma$ further the Taylor vortices become
compressed, cf., the wave number plot. Then the number of nodes of $u$ reduces
to 8 and finally to 6 as a vortex pair is annihilated in the center and then
yet another one. The compression prior to the vortex pair annihilation and the
relaxation to the old $k$-value after the annihilation can be seen in the top
plot of Fig.~\ref{fig6}.

\subsubsection{Increasing $\Gamma$}\label{SEC:decreaseGamma}

We also did a reverse ramp simulation in which the length was
increased by moving the top lid upwards from $\Gamma =5$ to $\Gamma
=16$ starting from TVF with very small admixtures of $m \neq 0$
modes as they are still present shortly after a start from rest. The
time intervals between upwards steps of $\Delta \Gamma=0.05$ was 2
radial diffusion times. This time interval, however, is not long
enough to allow for the full development of the spiral generating
defects that are described in detail in Sec.~\ref{SEC:rot-defect}.
Here the SPI flow permanently re-appeared in the center only at
$\Gamma \simeq 10$ whereas it had disappeared in the downwards ramp
at $\Gamma \simeq 8.3$. In addition we found in the ramp simulations
that the $\Gamma$ values at which the transitions from SPI to TVF
and vice versa occurred are affected also by the relative directions
of the lid motion and the SPI propagation. The reasons for this
hysteresis are on the one hand the upwards ramp being too fast but
also an inherent bistability between TVF and SPI flow in this small
system that is suggested by the following simulation:

\subsubsection{Different initial conditions}

\begin{figure}[ht]
\begin{center}
\includegraphics[width=12cm]{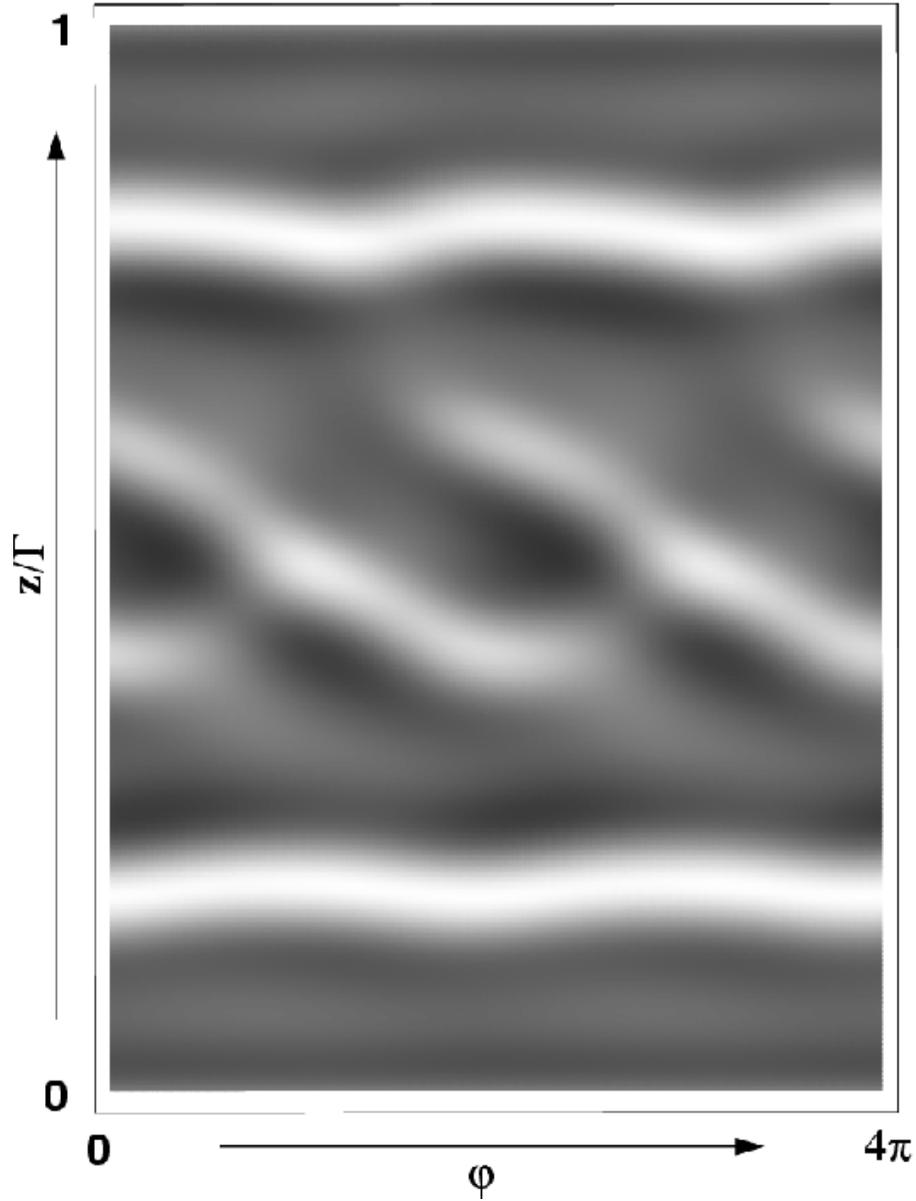}
\end{center}
\caption{Rotating defected vortex structure in a short system, $\Gamma=5.85$.
Gray scale plot shows the radial velocity field $u$ in an unrolled cylindrical
$\varphi$-$z$-surface at mid gap. White (black) denotes radial out (in)
flow. For better visibility the plot is periodically extended to an azimuthal
interval of $4\pi$. Parameters are $R_1=115, R_2=-100$.}
\label{fig7}
\end{figure}

\begin{figure}[ht]
\begin{center}
\includegraphics[width=12cm]{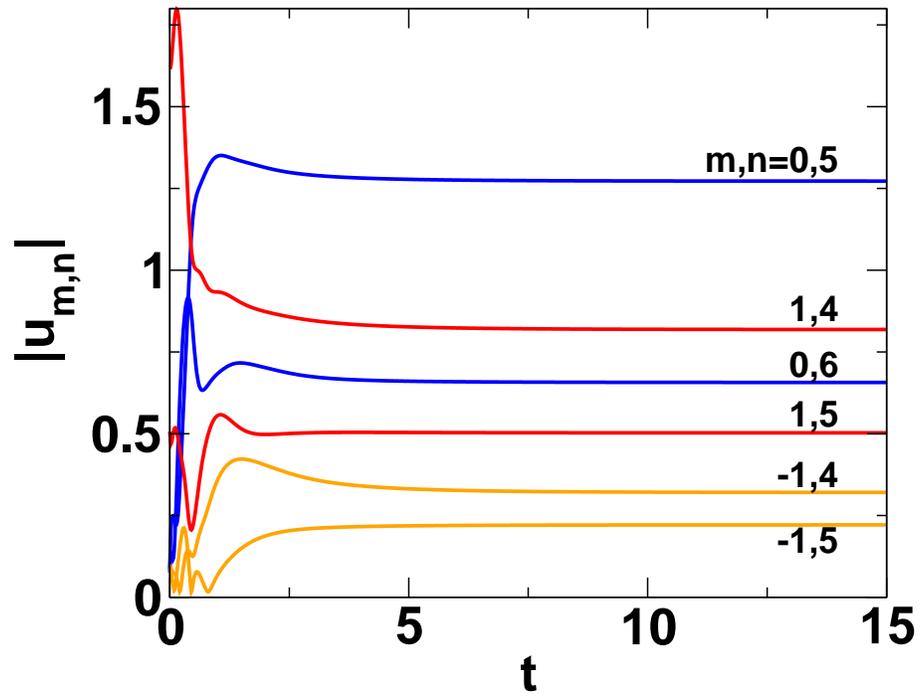}
\end{center}
\caption{ Time evolution towards the defected vortex state of Fig.~\ref{fig7}.
Shown are dominant mode amplitudes $|u_{m,n}|$
(\ref{FFT-expansion}) of the radial velocity field at mid gap. Here a perfectly
periodic L-SPI structure was instantaneously subjected to rigid-lid boundaries at
$z=0$ and $z=\Gamma=5.85$.}
\label{fig8}
\end{figure}

Here we started with a perfect, axially periodic L-SPI structure of
wavelength $\lambda=1.6$ at $R_1=115, R_2=-100$. Then we imposed
instantaneously
the rigid-lid boundary conditions at $z=0$ and $z=\Gamma=5.85$. Soon a defected
vortex structure appeared ( cf., Fig.~\ref{fig7}) that rotates as a whole
like a rigid body. But the phase
propagates axially upwards only in a very small central region where the white
stripes in Fig.~\ref{fig7} are tilted to the left. The time
evolution of the six largest mode amplitudes $|u_{m,n}|$ (\ref{FFT-expansion})
of the radial velocity field at mid gap towards this final state are shown in
Fig.~\ref{fig8}. So, this
vortex solution is dominated by the $m=0$ modes from the Ekman
vortices. Then it contains $m=1$ modes with L-SPI character but there is also a
significant admixture of $m=-1$ modes with R-SPI character. 

\subsubsection{Remarks}

Obviously the control parameter range in the $R_2-R_1$-plane of
Fig.~\ref{fig1} in which SPI are stable in finite length systems depends on
$\Gamma$.
Reducing $\Gamma$ will shrink the range of SPI flow eventually to zero because
of the ever present Ekman vortices in finite
length systems. In addition, the Ekman vortices prevent also
to reach the full stability domain of SPI under axially periodic conditions
when $\Gamma$ is increased. For example, at $R_2=-100$ we could not obtain
stable SPI flow for $R_1 \gtrsim 123$, i.e., in a domain where SPI solutions coexist
bistably with TVF solutions when axially periodic boundary conditions prevent
Ekman vortices.

We checked that our numerically obtained stability boundaries
largely agree with experimental ones \cite{SP00}. But
in the above described downwards-ramp-simulation we do not see SPI anymore
for $\Gamma<8.3$ and in particular not
at $\Gamma=5.85$ (where they are reported, e.g., for our $R_1=115, R_2=-100$
in Fig.~3 of Ref.~\cite{LPA04}) but rather TVF, i.e, a pure $M=0$ stationary
state. However, when starting from
different initial conditions with different histories we do see there SPI-like
phase propagation with several modes being present. Thus, there seems to be
multi- or at least bistability of pure $M=0$ vortex flow states coexisting with
mixed-mode ones.

We finally mention that the way how SPI flow in the center is destroyed or
generated
depends on the way the relevant parameters, say, $R_1$ and $\Gamma$
are varied. In Fig.~\ref{fig6} $\Gamma$ was decreased quasi-statically
causing a reduction of the SPI extension that was almost quasi-static except
for the last instance. What happened there can be better observed in a
different simulation: starting at $R_1=115$ with stable L-SPI in a long system
the inner Reynolds number is stepped up instantaneously into the instability
range of SPI flow, $R_1 \gtrsim $ 120. Then a fast TVF front
propagates upwards. It originates from the Ekman vortex structure and it
pushes the SPI phase generating defect upwards. The Ekman vortex
structure at the upper lid, however, is unable to trigger a downwards
propagating TVF front against the upwards traveling L-SPI phase. In fact the
phase annihilating defect below the upper Ekman vortex structure seems to
be more robust. Finally there could arise local wavy vortex flow at large
enough $R_1$ or TVF. But we have also observed for smaller $R_1$
counter propagating spirals which originate from a defect in the center.

\section{Conclusion}

We have numerically investigated how SPI flow is realized in finite length
Taylor Couette systems in which stationary top and bottom lids close the
annulus, i.e., in the presence of spatially localized Ekman vortices. Results
are presented for several system lengths $5 \leq \Gamma \leq 16$. In the
parameter range investigated here SPI solutions are stable under axially
periodic boundary conditions. But TVF solutions would be unstable there under these
idealized conditions without Ekman vortices. The presence of the latter in real
systems tends to stabilize TVF and to destabilize SPI flow.

For example, in a start-from-rest simulation with small initial noise
one can observe the following scenario: First, on a short time scale of 1-2
radial diffusion times, the unstable CCF is growing
radially in the bulk and simultaneously Ekman vortices are growing near the
lids. Then fast TVF fronts propagate axially into the bulk from the
Ekman vortex structures. Thereby unstable CCF is replaced by TVF
within a few diffusion times. For those parameters for which this TVF
is unstable in finite systems on can then observe a slow transformation of TVF
to SPI flow. Therein a pair of bulk TVF vortices becomes more and more deformed
and gets pinched together at a defect that 'cuts' them into two. They move apart,
get reoriented, and reconnect differently to form locally a spiral vortex pair.
This defect formation and reconnection is repeated at two new locations further
upwards and downwards towards the lids. Finally the axial defect propagation is
stopped by the strong Ekman vortex structures.

So, in the final state the bulk is filled with, say, an axially upwards
propagating L-SPI structure. Its phase is generated by a defect that is
rotating in the lower part of the system. The spiral phase is annihilated at
another rotating defect in the upper part of the system. The Ekman vortex
structures  are only slightly indented and modulated by the respective rotating
defect. The whole flow structure is rotating as a whole like a rigid body with
a global rotation rate into the same positive $\varphi$-direction as the inner
cylinder.

The SPI structure in the bulk is uniquely selected. Its wave number is
for large system lengths practically independent of $\Gamma$ showing a slight
variation only near the transition to TVF at small $\Gamma$.
When changing quasi-statically the system length at fixed $R_1, R_2$ the axial
extension over which SPI flow is realized in the bulk changes accordingly. The
Ekman vortex structures, on the other hand, remain basically unaffected. Below a
critical $\Gamma$ the two Ekman vortex structures have come too close to allow
for SPI flow any more.

It would be interesting to quantitatively test the frequency and
wave number selection in the SPI bulk, the structural dynamics of
the rotating defects, and the interpenetrating SPI and Ekman vortex
modes by spatiotemporal Fourier analyses of experimental data.

\section*{Acknowledgment} 
This work was supported by the Deutsche Forschungsgemeinschaft.

\clearpage




\begin{thebibliography}{99}
\bibitem{reviews}
For an account of the early history see, e.g., ref.~\cite{D91}. Later reviews
are contained in refs.~\cite{DPS85,CH93,T94,CI94}.
\end{thebibliography}
\end{document}